\documentclass[a4paper]{jpconf}
\usepackage{citesort}
\usepackage{graphicx}
\usepackage[colorinlistoftodos]{todonotes}
\usepackage[utf8]{inputenc} 
\begin{document}
\title{The fluctuations of quadrangular flow}
\author{Giuliano Giacalone$^1$, Li Yan$^1$, Jacquelyn Noronha-Hostler$^2$ and 
Jean-Yves Ollitrault$^1$}
\address{$^1$ Institut de physique th\'eorique, Universit\'e Paris Saclay, CNRS,
CEA, F-91191 Gif-sur-Yvette, France} 
\address{$^2$ Department of Physics, University of Houston, Houston TX 77204, USA}
\ead{jean-yves.ollitrault@cea.fr}

\begin{abstract}
The ATLAS Collaboration has measured for the first time the fourth
cumulant of quadrangular flow, $v_4\{4\}^4$. Unlike the fourth cumulants of 
elliptic and triangular flows, it presents a change of sign above 30\% centrality. 
We show that this change of sign is predicted by event-by-event
hydrodynamics. We argue that it results from the combined effects of
a nonlinear hydrodynamic response, which couples quadrangular flow
to elliptic flow, and elliptic flow fluctuations. 
\end{abstract}

\section{Introduction}
The bulk of particle production in ultrarelativistic nucleus-nucleus
collisions is described by the flow paradigm~\cite{Luzum:2011mm},
which states that particles are emitted independently from an
underlying probability distribution. 
In particular, the flow paradigm naturally explains
the long-range azimuthal correlations, which are a salient feature of
heavy-ion collisions, as resulting from 
the fluctuations of the underlying azimuthal probability
distribution $P(\varphi)$~\cite{Alver:2010gr}. This azimuthal distribution is
traditionally written as a Fourier series:
\begin{equation}
\label{defVn}
P(\varphi)=\frac{1}{2\pi}\sum_{n=-\infty}^{+\infty}V_n \textup{e}^{-in\varphi},
\end{equation}
where $V_n=v_n\exp(in\Psi_n)$ 
is the (complex) anisotropic flow coefficient in the $n$th
harmonic and $V_{-n}=V_n^*$. 
Both the magnitude and phase of $V_n$ fluctuate event to
event~\cite{Alver:2006wh}. 
Experimental observables involving anisotropic flow can be recast as
statistical properties of the distribution of $V_n$. 
For instance, the cumulants $v_n\{2\}^2$ and $v_n\{4\}^4$ are defined 
by~\cite{Borghini:2001vi}:
\begin{eqnarray}
\label{defvn24}
v_n\{2\}^2&\equiv&\langle v_n^2\rangle,\cr
v_n\{4\}^4&\equiv&2\langle v_n^2\rangle^2-\langle v_n^4\rangle,
\end{eqnarray}
where angular brackets denote an average over events in a centrality
class. 
The lowest order cumulant, $v_n\{2\}^2$, is simply the mean square value of $v_n$,
while the fourth cumulant, $v_n\{4\}^4$, is a nontrivial combination of 
moments. 
In particular, despite the unfortunate notation, $v_n\{4\}^4$ can be
either positive or negative, depending on the probability distribution of $v_n$. 
Both $v_2\{4\}^4$~\cite{Aamodt:2010pa} 
and $v_3\{4\}^4$~\cite{ALICE:2011ab} are observed to be positive across
all centralities in Pb+Pb collisions at the LHC. 
However, a striking observation, which seems to have received little
attention from the theory community, is that $v_4\{4\}^4$ changes
sign~\cite{Aad:2014vba}: it is positive only up to $\sim 30\%$ centrality. 

In this article, we show that the change of sign of $v_4\{4\}^4$ is
predicted by hydrodynamics. 
In hydrodynamics, the fluctuations of the anisotropic flow coefficients, $V_n$, result from the fluctuations of the energy density profile released after the collisions~\cite{Teaney:2010vd}.
$v_2$ and $v_3$ are determined to a good
approximation~\cite{Niemi:2012aj,Gardim:2014tya} by linear
response to the initial anisotropies in the corresponding harmonics. 
This, in turn, explains why both $v_2\{4\}^4$ and $v_3\{4\}^4$ are
positive, even though they originate from different mechanisms: 
$v_2\{4\}$ is driven by the reaction-plane eccentricity~\cite{Voloshin:2007pc}
while $v_3\{4\}$ is driven by non-Gaussian fluctuations of the initial triangularity~\cite{Bhalerao:2011bp,Yan:2013laa,Gronqvist:2016hym}.
By contrast, simple linear response to the initial anisotropy in the
fourth harmonic is unable to explain the observed fluctuations of
$v_4$~\cite{Rybczynski:2015wva,Ghosh:2016npo}. 
In Sec.~\ref{s:response}, we recall why 
linear response does not apply to
$v_4$ and how a significant nonlinear response can be taken into
account~\cite{Gardim:2011xv}. 
We infer the nonlinear response from
experimental data and we show that its magnitude is correctly predicted
by hydrodynamics. 
In Sec.~\ref{s:v44}, we calculate $v_4\{4\}^4$ in hydrodynamics. 

\section{Linear and nonlinear hydrodynamic response}
\label{s:response}

$V_4$ and $(V_2)^2$ transform identically under azimuthal
rotations. Therefore, rotational symmetry allows for a coupling between
these two quantities, which is indeed predicted by
hydrodynamics~\cite{Borghini:2005kd}. 
We take this coupling into account by writing $V_4$ as the sum of a
term proportional to $(V_2)^2$ (the nonlinear response) and a term uncorrelated with $(V_2)^2$, which we dub the linear part,
$V_{4L}$~\cite{Yan:2015jma,Giacalone:2016afq}:
\begin{equation}
\label{decomposition}
V_4=V_{4L}+\chi_4 (V_2)^2.
\end{equation} 
The condition that linear and nonlinear parts are uncorrelated,
$\langle V_{4L}(V_2)^{*2}\rangle=0$, 
uniquely defines the proportionality coefficient, $\chi_4$, i.e., 
\begin{equation}
\label{defchi4}
\chi_4=\frac{\langle V_4(V_2^*)^2\rangle}{\langle |V_2|^4\rangle}.
\end{equation} 
Note that the decomposition defined in Eqs.~(\ref{decomposition}) and
(\ref{defchi4}) is purely mathematical and holds 
irrespective of the hydrodynamical 
setup.\footnote{As an example, hydrodynamics predicts a similar
  decomposition~\cite{Teaney:2012ke}, 
where the linear part results from initial fluctuations in the fourth
harmonic. In this case, there is no requirement concerning how the linear and the nonlinear part are correlated.} 
The nonlinear part thus defined can be isolated by analyzing $V_4$
with respect to 
the direction of elliptic flow~\cite{Adams:2003zg,Adare:2010ux}. The
resulting observable is dubbed $v_4\{\Psi_2\}$~\cite{Gardim:2012yp}
and is defined by~\cite{Luzum:2012da}:
\begin{equation}
\label{v4psi2}
v_4\{\Psi_2\}\equiv\frac{\langle V_4(V_2^*)^2\rangle}{\sqrt{\langle
    |V_2|^4\rangle}}.
\end{equation}
From Eqs.~(\ref{defchi4}) and (\ref{v4psi2}), one obtains 
$v_4\{\Psi_2\}=\chi_4\sqrt{\langle |V_2|^4\rangle}$, that is, 
$v_4\{\Psi_2\}$ is the rms value of the nonlinear contribution to
$V_4$ in Eq.~(\ref{decomposition}). 
The triangular inequality guarantees that
$v_4\{\Psi_2\}\le v_4\{2\}$~\cite{Giacalone:2016afq}, where $v_4\{2\}$ is defined in 
Eq.~(\ref{defvn24}). The mean square value of $V_{4L}$ can be obtained by
subtracting $v_4\{\Psi_2\}^2$ from
$v_4\{2\}^2$~\cite{Jia:2014jca,Qian:2016pau}. 
\begin{figure}[ht]
\begin{minipage}{.47\linewidth}
\includegraphics[width=\linewidth]{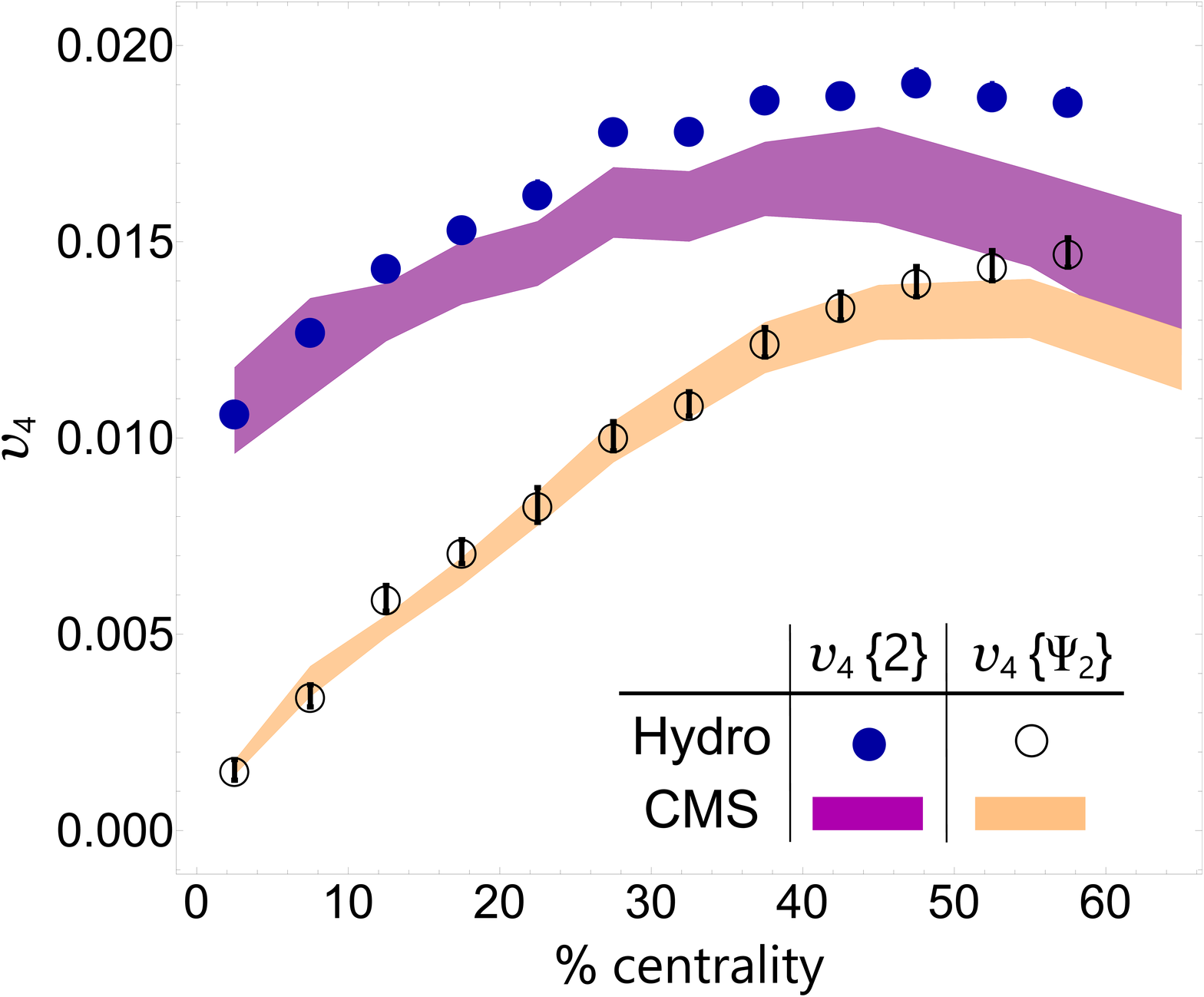}
\caption{\label{fig:v42}$v_4\{2\}$ (full symbols and dark shaded band)  and $v_4\{\Psi_2\}$
(open symbols and light shaded band)
as a function of centrality percentile in Pb+Pb collisions at 2.76~TeV. 
Bands: CMS data for charged particles in the $p_t$ range $0.3<p_t<3$~GeV~\cite{Chatrchyan:2013kba}.  
Symbols: hydrodynamic calculations (pions, same $p_t$ range). }
\end{minipage}\hspace{.06\linewidth}%
\begin{minipage}{.47\linewidth}
\includegraphics[width=\linewidth]{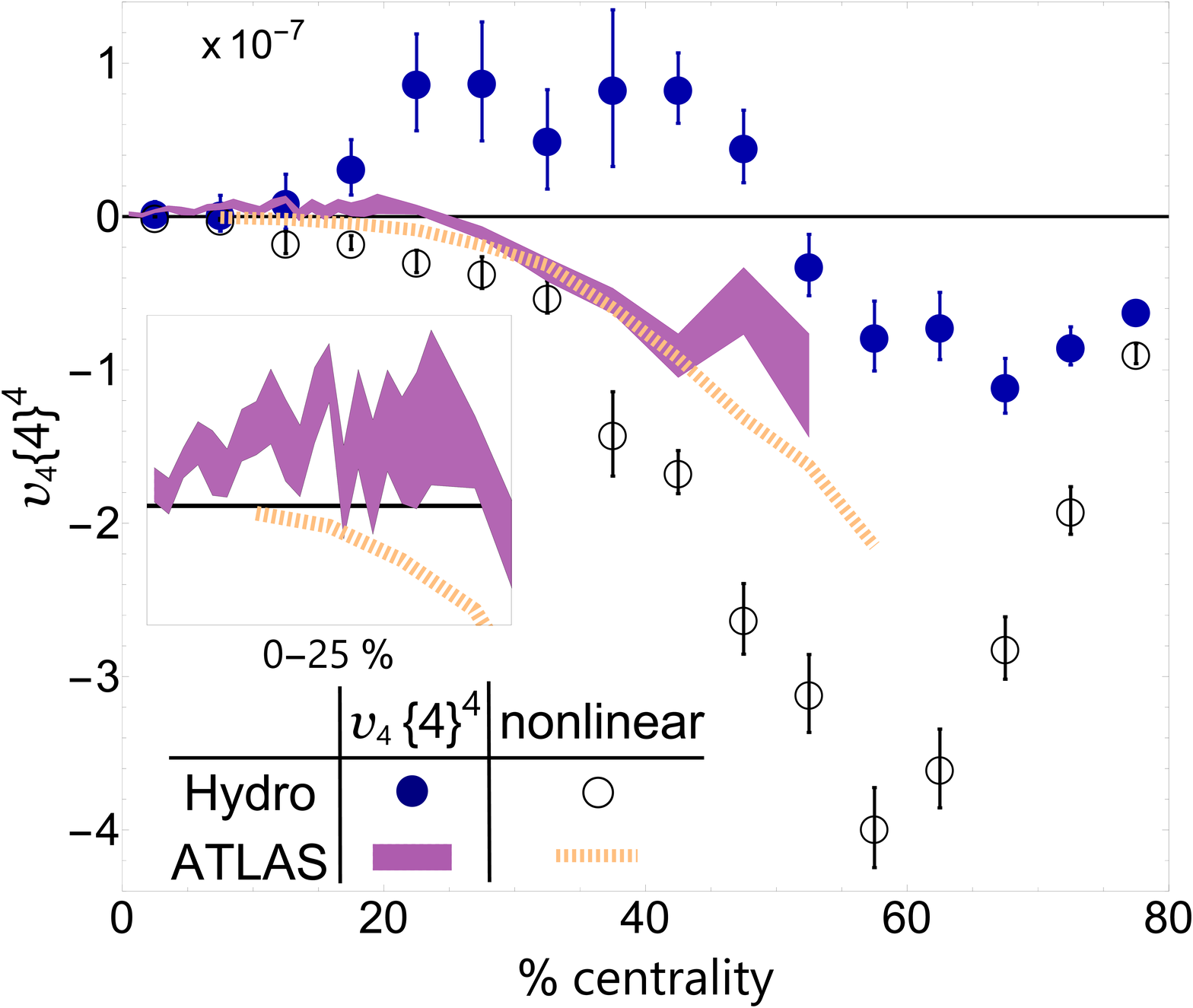}
\caption{\label{fig:v44}
Dark shaded band: ATLAS data~\cite{Aad:2014vba} for $v_4\{4\}^4$ as a function
of centrality percentile for charged particles in the $p_t$ range
$0.5<p_t<5$~GeV. 
Full symbols: hydrodynamic calculation (pions, $0.5<p_t<3$~GeV). 
Dashed line and open symbols: last term of Eq.~(\ref{nlpart}), 
from ATLAS data and hydrodynamic calculations. 
}

\end{minipage} 
\end{figure}
Figure~\ref{fig:v42} displays CMS data~\cite{Chatrchyan:2013kba}  for
$v_4\{\Psi_2\}$ and $v_4\{2\}$. They satisfy 
$v_4\{\Psi_2\}<v_4\{2\}$ for all centralities, which is a nontrivial
test of the flow paradigm. 
Further, because of the increase of elliptic flow in the reaction plane, the relative weight of the nonlinear contribution, $v_4\{\Psi_2\}$, increases with centrality. 
In order to illustrate that these trends are captured by
hydrodynamics~\cite{Qiu:2012uy,Teaney:2013dta,Niemi:2015qia}, 
we carry out event-by-event viscous relativistic hydrodynamic
calculations within the code v-USPhydro
\cite{Noronha-Hostler:2013gga,Noronha-Hostler:2014dqa}. 
Initial conditions are given by the Monte Carlo Glauber
model~\cite{Alver:2008aq}.  
The setup is the same as in Ref.~\cite{Noronha-Hostler:2015dbi}: In
particular, the shear viscosity over entropy ratio is
$\eta/s=0.08$~\cite{Policastro:2001yc}. 
Anisotropic flow, $V_n$,  is calculated at 
freeze-out~\cite{Teaney:2003kp} for pions in every event.
We calculate both $v_4\{2\}$ and $v_4\{\Psi_2\}$ by averaging over
events. Results are shown in Fig.~\ref{fig:v42}.
Our calculation matches experimental data on 
$v_4\{\Psi_2\}$ but slightly overestimates $v_4\{2\}$, meaning
that our hydrodynamical setup overestimates the linear part, $V_{4L}$. 
We stress that, in our calculation, we implement a very low value of
$\eta/s$ and that the linear part, $V_{4L}$, is more strongly damped
by viscosity than the nonlinear part~\cite{Teaney:2012ke}. Agreement
with data is likely to be  improved with a larger value of $\eta/s$. 

\section{Explaining $v_4\{4\}^4$}
\label{s:v44}

Figure~\ref{fig:v44} presents $v_4\{4\}^4$, as 
defined in Eq.~(\ref{defvn24}), from ATLAS data~\cite{Aad:2014vba} 
(the plotted quantity is $c_4\{4\}\equiv-v_4\{4\}^4$).
It is positive up to 25\% centrality (see inset in Fig.~\ref{fig:v44}) and then negative. 
This change of sign is also observed in hydrodynamic calculations (full symbols), 
although it occurs around $50\%$ centrality.  
We now argue that the change of sign is driven by
the nonlinear response. Neglecting the linear part, $V_{4L}$, in Eq.~(\ref{decomposition}), one obtains
\begin{equation}
\label{nlpart}
v_4\{4\}^4=\chi_4^4\left(2\langle v_2^4\rangle^2-\langle
v_2^8\rangle\right)=
v_4\{\Psi_2\}^4 \left(2-\frac{\langle v_2^8\rangle}{\langle
  v_2^4\rangle^2} \right),
\end{equation}
where, in the last equality, we have used Eqs.~(\ref{defchi4}) and
(\ref{v4psi2}). 
We compute the last term of Eq.~(\ref{nlpart}) both in hydrodynamics and using experimental data.
The estimate of hydrodynamics is shown as open symbols in Fig.~\ref{fig:v44}.
As for experimental data, we employ the relation $v_4\{\Psi_2\}\equiv v_4\{2\}\langle\cos(4(\Phi_4-\Phi_2))\rangle_w$, where $\langle\cos(4(\Phi_4-\Phi_2))\rangle_w$ is the event-plane correlation ~\cite{Yan:2015jma,Aad:2014fla} measured by the ATLAS Collaboration.
Higher-order moments of $v_2$ are instead obtained from the measured higher-order cumulants of elliptic flow~\cite{Aad:2014vba,Yan:2015jma}.
The resulting estimate is plotted as a dashed line in Fig.~\ref{fig:v44}.
Both estimates show that large fluctuations of $v_2$ lift the value of $\langle v_2^8 \rangle / \langle v_2^4 \rangle^2$, causing the contribution of the nonlinear term to be negative for all centralities. The nonlinear term increases in magnitude as a
function of centrality percentile and drives the change of sign of $v_4\{4\}^4$.
We find that the hydrodynamic calculation overestimates both $v_4\{4\}^4$ and the 
difference between $v_4\{4\}^4$ and the nonlinear contribution.
Moreover, the change of sign of $v_4\{4\}^4$ occurs at a centrality percentile which is too large.
These issues are consistent with the conclusion drawn in 
Fig.~\ref{fig:v42}:  Our hydrodynamical setup overestimates the linear part, $V_{4L}$. 

We have shown that the peculiar centrality dependence of $v_4\{4\}^4$
observed in Pb+Pb collisions at the LHC is understood in
hydrodynamics as resulting from the combined effects of a
nonlinear hydrodynamic response coupling $v_4$ to $v_2$, and large
$v_2$ fluctuations. 
This provides further evidence for a fluidlike behavior of the matter created in
ultrarelativistic Pb+Pb collisions. 

\ack
This work is 
supported by the European Research Council under the
Advanced Investigator Grant ERC-AD-267258. 
JNH~was supported by the National Science Foundation under grant no.~PHY-1513864
and underneath FAPESP grant: 2016/03274-2.
The authors wish to thank the ATLAS Collaboration for providing experimental data on $c_4\{4\}$.  
\section*{References}
\bibliography{v4bib}

\end{document}